\documentclass[journal]{IEEEtran}
\usepackage[thinlines]{easytable}
\usepackage{mathtools,amssymb,bm}
\usepackage{amsmath}
\usepackage{cite}
\usepackage{graphicx}
\usepackage{epstopdf}

\usepackage{cite}
\newtheorem{assumption}{Assumption}
\usepackage[utf8]{inputenc}
\usepackage{longtable}
\usepackage[ruled,vlined]{algorithm2e}
\usepackage{color}
\usepackage{caption}
\usepackage{multirow}
\usepackage[footnotesize]{subfigure}
\usepackage{hyperref}
\newtheorem{thm}{\textbf{Theorem}}

\begin{document}
	\title{ ISAC Super-Resolution Receivers: The Effect of Different Dictionary Matrices}
	\author{Iman Valiulahi, Christos Masouros, \textit{Fellow,} IEEE, {\color{black} Athina P. Petropulu, \textit{Fellow}, IEEE} 
	\thanks{Tman Valiulahi and Christos Masouros are with the Department of Electronic and Electrical Engineering, University College London, London WC1E 7JE, U.K. (e-mails: i.valiulahi@ucl.ac.uk; c.masouros@ucl.ac.uk). {\color{black} Athina P. Petropulu is with the Department of
		Electrical and Computer Engineering, Rutgers, The State University of New
		Jersey, Piscataway, NJ 08854 USA (email:
		athinap@soe.rutgers.edu).} }	}
	
	\maketitle

	\begin{abstract}
		This paper presents an off-the-grid estimator for ISAC systems using lifted atomic norm minimization (LANM). The main challenge in the ISAC systems is the unknown nature of both transmitted signals and radar-communication channels. We use  a known dictionary to encode transmit signals and show that LANM can localize radar targets and decode communication symbols when the number of observations is proportional to the system's degrees of freedom and the coherence of the dictionary matrix. We reformulate LANM using a dual method and solve it with semidefinite relaxation (SDR) for different dictionary matrices to reduce the number of observations required at the receiver. Simulations demonstrate that the proposed LANM accurately estimates communication data and target parameters under varying complexity by selecting different dictionary matrices.
	\end{abstract}
	
	
	\section{Introduction}
Recent interest in communication and radar spectrum sharing (CRSS) has led to two main approaches: radar-communication coexistence and integrated sensing and communication (ISAC) systems \cite{liu2020joint, meng2023network}. Radar-communication coexistence develops interference management strategies, allowing both systems to operate without disrupting each other \cite{vargas2023dual}. ISAC integrates sensing and communication within the same system, providing real-time cooperation \cite{valiulahi2023net,alaaeldin2023robust,valiulahi2022antenna,virgili2022cost}.

ISAC systems not only boost spectral and energy efficiency but also lower hardware and signaling costs by combining both functions \cite{liu2022integrated, hua1993pencil}. These advantages have expanded ISAC's applications to fields such as vehicular networks, indoor positioning, and covert communications \cite{meng2023sensing, meng2024cooperative, blunt2010intrapulse}. Unlike traditional approaches that consider sensing and communication as separate systems, ISAC mutually designs them for mutual benefits, maximizing resource use and operational effectiveness \cite{liu2018toward,han2024beamspace}.

The main challenge in ISAC reception is that both transmit signals and channels are unknown, making traditional radar and communication methods ineffective. ISAC receivers estimate both target parameters and communication data simultaneously \cite{liu2020joint,hu2024isac}. Techniques like MUSIC \cite{schmidt1986multiple} and ESPRIT \cite{yan2019two} struggle with noise and correlated targets, while compressed sensing (CS) improves resolution but assumes grid-based DOAs, leading to errors \cite{han2022high, chi2011sensitivity}. These methods also fail to jointly detect DOAs and communication data.

To address basis mismatch, atomic norm minimization (ANM) was developed, promoting signal sparsity in a continuous dictionary, serving as a continuous version of $\ell_{1}$ norm minimization. ANM has been applied in MIMO radar \cite{tang2020range}, line spectral estimation \cite{valiulahi2019robustness}, and OFDM noise elimination \cite{valiulahi2019eliminating}, outperforming conventional methods. It has also been used for delay-Doppler and angle-delay-Doppler pair recovery \cite{heckel2016super, heckel2016super1}. Lifted ANM (LANM) handles blind detection in super-resolution problems by estimating both spectral coefficients and transmit signals \cite{chi2016guaranteed} \cite{bigdeli2022noncoherent}, addressing AWGN and differentiating radar from communication signals in spectrum coexistence \cite{suliman2021mathematical, vargas2023dual}.
 We propose a novel ISAC receiver design based on LANM that can simultaneously recover target locations, velocities, delays, DOAs, and communication data from reflected signals. Despite the previous developments on the LANM in one dimensional signaling, we introduce a theorem that adapts the LANM technique for MIMO scenarios in ISAC systems.

\begin{figure}
		\centering
		\includegraphics[width=1\linewidth]{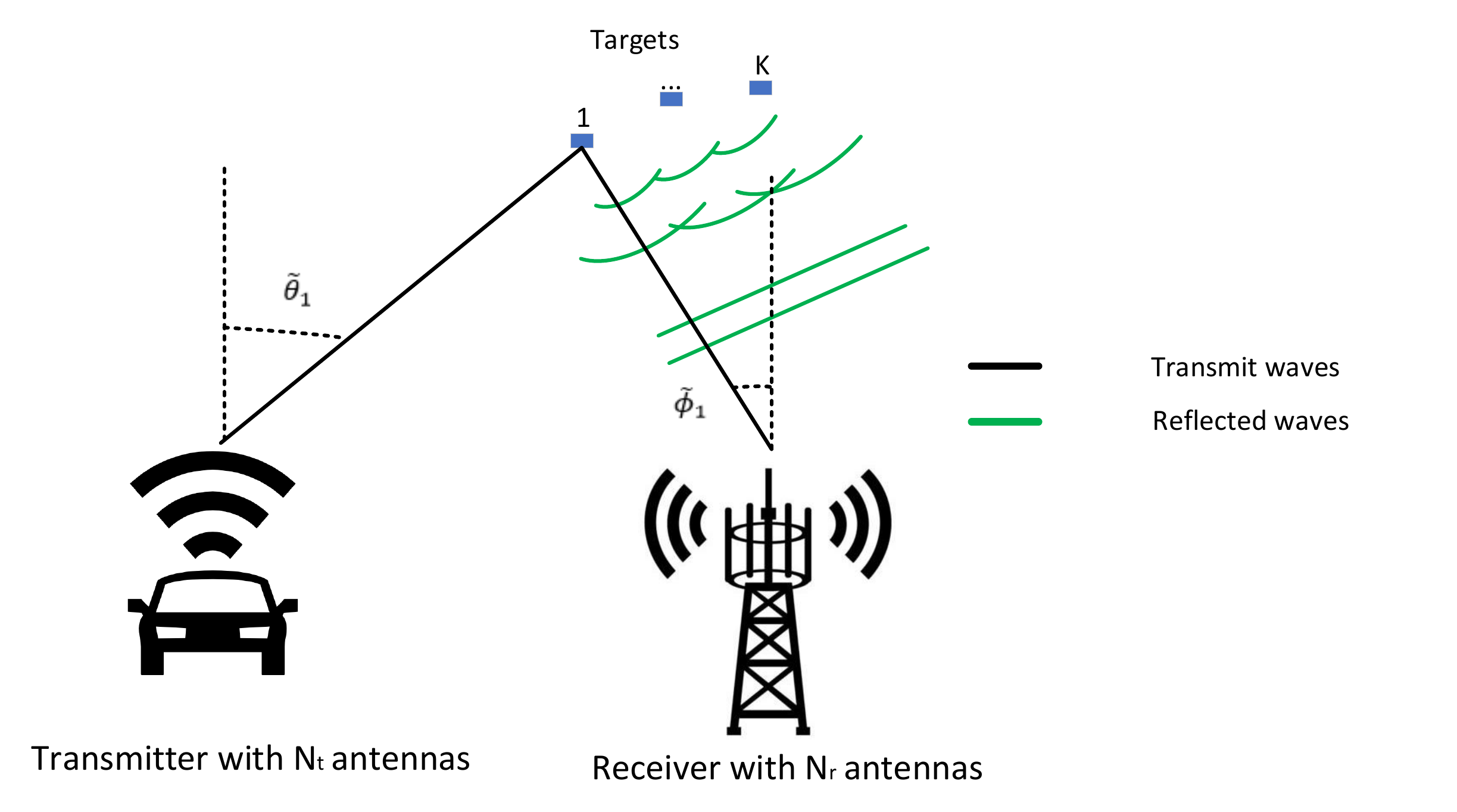}
		\caption{System model.}
		\label{fig:drawing1}
	\end{figure}

 Consider a bistatic radar scenario as shown in Fig. \ref{fig:drawing1}. The transmitter illuminates signals while the receiver collects echoes from $K$ targets. The challenge is to simultaneously detect the target parameters and estimate the transmitted signals with communication data. This problem is bilinear and difficult to solve. Traditionally, it is assumed that a reference link or pilot signals are used, which wastes bandwidth. The estimated signal may have demodulation errors, leading to reduced performance \cite{zheng2017super}. Additionally, physical obstructions might block the radio waves, making the direct link unavailable.

We propose a LANM-based approach, assuming transmitted signals lie on a known low-dimensional random dictionary. The received signal is modeled as a sparse combination of low-rank matrices, addressed using LANM. We derive a dual formulation to ensure exact solutions under certain conditions and prove that the number of samples required for accurate estimation is proportional to the number of targets, transmit symbols, and the coherence of the compression matrix. Different compression matrices lead to varying receiver complexity, allowing us to reduce implementation complexity by either reducing the number of antennas or lowering the sampling rate. Simulations confirm that our estimator accurately detects both communication data and target parameters with adjustable complexity at the receiver.

	Here, we introduce the notation used in this paper.
	Vectors and matrices are denoted by boldface lowercase and uppercase letters, respectively and scalars or entries are non-bold lowercase. The $\|\cdot\|_1$ and $\|\cdot\|_2$ are $\ell_1$ and $\ell_2$ norms, respectively. The operators $\mathrm{tr}(\cdot)$ and  $(\cdot)^H$ are trace of a matrix, hermitian of a vector, respectively.

	\section{System Model and Problem Formulation}\label{systems}
	We consider an ISAC system with a transmitter having $N_t$ antennas and a receiver with $N_r$ antennas, targeting $K$ objects, where $K < N_t$. The targets are located in the far field of the arrays, with uniformly spaced transmit and receive antennas, where the spacing is $\frac{N_t}{2f_c}$ and $\frac{N_r}{2f_c}$, respectively. The baseband received signal $y_r(t)$ at the $r$-th receive antenna, $r = \{0, \cdots, N_r-1\}$, is the sum of reflected signals from the targets, denoted by $x_k(t)$ for $k = \{0, \cdots, K-1\}$. Mathematically, this can be expressed as:
	\begin{align}\label{1}
		y_r(t) = \sum_{k=0}^{K-1} \sum_{s=0}^{N_t-1} \alpha_k e^{i2\pi r N_t \phi_k} e^{i2\pi s \theta_k} x_k(t - \bar{\tau}_k) e^{i2\pi \bar{v}_k t},
	\end{align}
	where $i=\sqrt{-1}$, $\alpha_k \in \mathbb{C}$ is the complex attenuation factor representing radar path loss and cross-section, $\theta_k = -\frac{\sin(\tilde{\theta}_k)}{2}$ and $\phi_k = -\frac{\sin(\tilde{\phi}_k)}{2}$ are the angle of departure (AoD) and angle of arrival (AoA) for the $k$-th target, respectively, as shown in Fig. \ref{fig:drawing1}. $\bar{\tau}_k$ and $\bar{v}_k$ represent the delay and Doppler shifts for the $k$-th target. These parameters $\theta_k, \phi_k, \bar{\tau}_k, \bar{v}_k$ can be translated into the angle, distance, and velocity of each target relative to the radar. 
	Our goal is to recover the parameters $\alpha_k, \theta_k, \phi_k, \bar{\tau}_k, \bar{v}_k$ from the received signal $y_r(t)$ and estimate the probing signals $x_k(t)$ containing communication data. There is no assumed direct transmitter-receiver link, as this could waste bandwidth and introduce demodulation errors \cite{zheng2017super}, and obstacles may block direct paths. The transmit signal is assumed to be band- and time-limited, and the received signal is sampled at a rate of $\frac{1}{B}$ over the interval $[0, T_t]$, where the samples are collected into a vector $\bm{y}_r \in \mathbb{C}^{\bar{L}}$, with $\bar{L}:= BT$. The $p$-th entry of $\bm{y}_r$ is given by:
	\begin{align}\label{2}
		[\bm{y}_r]_p &= \frac{1}{\bar{L}} \sum_{k=0}^{K-1} \sum_{s=0}^{N_t-1} |\alpha_k| e^{i2\pi r N_t \phi_k} e^{i2\pi (s \theta_k +\theta^{\alpha}_{k})} \nonumber \\
		& \sum_{r=-N}^{N} \Bigg[\Bigg( \sum_{l=-N}^{N} x_k(l) e^{-i2\pi\frac{rl}{\bar{L}}} \Bigg) e^{-i2\pi p \tau_k} \Bigg] e^{i2\pi \frac{rp}{\bar{L}}} e^{i2\pi v p},
	\end{align}
	where $\tau_k = \frac{\bar{\tau}_k}{T}$ and $v_k = \frac{\bar{v}_k}{B}$ are the normalized time and frequency shifts, respectively. Note that we can write $\alpha_{k}=|\alpha_{k}|e^{i2\pi\theta_{k}^{\alpha}}$. The goal is now to estimate the locations of the targets, which requires determining the parameters $\alpha_k$, $(\theta_k, \phi_k, \tau_k, v_k) \in [0,1)^4$, and recovering the probing signal $\bm{x}_k$, $k = \{0, \cdots, K-1\}$ that contains communication data, from the observation model in (\ref{2}).	
	The number of unknowns in (\ref{2}) is in order of $\mathcal{O}(\bar{L}S)$, far exceeding the number of observations, making the problem ill-posed. 
	
	To address this, we assume the probing signals $\bm{x}_k \in \mathbb{C}^{\bar{L} \times 1}, k = \{0, \cdots, K-1\}$, lie in a known low-dimensional subspace, represented by a matrix $\bm{D} \in \mathbb{C}^{\bar{L} \times T}$ with $T \ll \bar{L}$, which we call it the compression matrix. Hence, $\bm{x}_k = \bm{D} \bm{h}_k$, where $\bm{h}_k \in \mathbb{C}^{T \times 1}$ is an unknown vector carrying the communication data. Recovering $\bm{x}_k$ reduces to estimating $\bm{h}_k$, as $\bm{D}$ is known by both transmitter and receiver.  Without loss of generality, we assume
	that $\|\bm{h}_{k}\|_{2}= 1$ for all $k$. Note that recovering $\bm{x}_{k}$ is equivalent to
	estimating $\bm{h}_{k}$ since  $\bm{D}$ is known by both the transmitter and receiver.

	Define $\bm{y}=[\bm{y}_0, \cdots, \bm{y}_{N_r-1}]^{T}$ and $\bm{\tau}_k := [\theta_k, \phi_k, \tau_k, v_k]^{T}$. The array response $\bm{a}(\bm{\tau}_k)$ is defined as:  
	\begin{align}
		[\bm{a}(\bm{\tau}_k)]_{(r, s, l, k, 1)} = e^{i2\pi r N_t \phi_k} e^{i2\pi (s \theta_k+\theta_{k}^{\alpha})} D_N&\bigg(\frac{l}{L}-\tau_k\bigg) \nonumber\\
		& D_N\bigg(\frac{r}{L}-v_k\bigg),
	\end{align}
	where $r \in \{0, \cdots, N_r-1\}$, $s \in \{0, \cdots, N_t-1\}$, $l,k \in \{-N, \cdots, N\}$, and $D_N(t)$ is the Dirichlet kernel. Using this, we rewrite (\ref{2}) as: 
	\begin{align}
		[\bm{y}]_j = \sum_{k=0}^{K-1} \alpha_k [\bm{a}(\bm{\tau}_k)]_{(r,s,l,k,1)} \bm{d}_{(j-l)}^H \bm{h}_k e^{i2\pi \frac{pr}{L}}.
	\end{align}
	With the linear operator $\mathcal{X}$ and its adjoint $\mathcal{X}^\star$, then, we can write
	\begin{align}\label{eq9}
		\bm{y} = \mathcal{X}(\bm{U}).
	\end{align}

Equation (\ref{eq9}) shows that recovery depends on $\bm{U}$, which is a sparse combination of matrices from the set of atoms
\begin{align}
	\mathcal{A} = \{\bm{v}\bm{a}(\bm{\tau})^{H} \mid \bm{\tau} \in [0,1)^{4}, \|\bm{v}\|_2 = 1\}.
\end{align}
The atomic norm of $\bm{U}$ is defined as the gauge function of $\mathrm{conv}(\mathcal{A})$:
\begin{align}
	\|\bm{U}\|_{\mathcal{A}} = \inf\Big\{\sum_k \alpha_k \mid \bm{U} = \sum_{k=0}^{K-1} \alpha_k \bm{h}_k \bm{a}(\bm{\tau}_k)^{H}, \alpha_k \geq 0 \Big\}.
\end{align}
To estimate $\mathcal{X}(\bm{U})$ from (\ref{eq9}), we propose minimizing the atomic norm:
\begin{align}
\min_{\bm{U}}~~& \|\bm{U}\|_{\mathcal{A}}\nonumber\\
	&\text{s.t.}~\bm{y}=\mathcal{X}(\bm{U}),
\end{align}
Solving this is difficult, so we consider its dual as below
\begin{align}
	\max_{\bm{q}} ~\langle\bm{q},\bm{y}_{w}\rangle_{\mathbb{R}} \quad \text{s.t.} \quad \|\mathcal{X}^{\star}(\bm{q})\|_{\mathcal{A}}^{\star} \leq 1,
\end{align}
where $\bm{q}$ is the dual variable and $\|\cdot\|_{\mathcal{A}}^{\star}$ is the dual atomic norm. In the following, we present the main results and theorem.

\begin{assumption}\label{assum1}
	We assume the columns of $\bm{D}^H$, denoted as $\bm{d}_{l} \in \mathbb{C}^{T\times 1}$, are selected independently and identically from a population $\mathcal{F}$, satisfying the following conditions:
	\begin{align}
		&\mathbb{E}[\bm{d}_{l}\bm{d}_{l}^{H}]=\bm{I}_{T}, ~ l=\{-N, \cdots, N\}, \nonumber\\
		&\max|d(i)|^{2} \le \mu,
	\end{align}
	where $\mu$ is the coherence parameter and $\bm{I}_{T}$ is the $T\times T$ identity matrix. We also assume $\mu T\geq 1$ for simplicity, achievable by selecting a sufficiently large $\mu$.
\end{assumption}

\begin{assumption}\label{assumption2}
	We assume all $\bm{h}_{k}$ are randomly selected from the complex unit sphere, i.e., $\|\bm{h}_{k}\|_{2} = 1$.
\end{assumption}

\begin{assumption}\label{assumption3}
	The radar parameters $(\tau_{k}, v_{k}, \theta_{k}, \phi_{k})$, $k =\{1, \cdots, K\}$, must satisfy the separation condition:
	\begin{align}
		\min_{k \neq k'} \max \bigg(& |\tau_{k}-\tau_{k^{\prime}}|,|v_{k}-v_{k^{\prime}}|,\nonumber\\&|\theta_{k}-\theta_{k^{\prime}}|,|\phi_{k}-\phi_{k^{\prime}}|\bigg)\geq \frac{10}{N_{t}N_{r}-1}
	\end{align}
	where $|a-b|$ is the wrap-around distance on the unit circle.
\end{assumption}
Note that $\bm{D}$, known at both transmitter and receiver, acts as a compression matrix, while $\bm{h}_{k}$ can encode different QAM signals for communication. Normalizing $\bm{h}_{k}$ still conveys QAM information, as the receiver knows the modulation scheme and signal size.
 Regarding the above discussion, we now present  our main theorem as follows:
 
  \begin{thm}\label{maintheorem}
Theorem \ref{maintheorem} states that for the linear system in (\ref{1}) and its sampled version in (\ref{2}), if the unknown waveforms can be expressed as $\bm{x}_{k} = \bm{D}\bm{h}_{k}$, where $\bm{D}$ satisfies Assumption \ref{assum1} and $\bm{h}_{k}$ follows Assumption \ref{assumption2}, and if the shifts meet the minimum separation in Assumption \ref{assumption3}, then with probability at least $1 - \delta$, the condition 
$
L^4 \geq C\mu KT \log\left(\frac{10KT}{\delta}\right)
$
ensures that $\bm{U}$ can be recovered through problem (\ref{eq9}). This result shows that, with enough radar samples relative to a logarithmic factor, an exact solution exists under mild separation conditions.
\end{thm}
 In the following, we generalize the proposed problem when the signal is corrupted by AWGN noise. Thus, we consider the following observation model 
\begin{align}
	\bm{y}_{w}=\mathcal{X}(\bm{U})+\bm{w},
\end{align}
containing both AWGN with the power $\|\bm{w}\|_{2} \le \sigma^{2}$. To estimate $\mathcal{X}(\bm{U})$ from (\ref{eq9}), we propose minimizing the atomic norm:
\begin{align}
	\min_{\bm{U}}~~& \|\bm{U}\|_{\mathcal{A}}\nonumber\\
	&\text{s.t.}~\|\bm{y}_{w}-\mathcal{X}(\bm{U})\|_{2}\le 
	\sigma^{2},
\end{align}
Consequently, we can write
\begin{align}\label{eq14}
	\max_{\bm{q}} ~\langle\bm{q},\bm{y}_{w}\rangle_{\mathbb{R}}-\frac{\sigma}{4}\|\bm{q}\|_{2} \quad \text{s.t.} \quad \|\mathcal{X}^{\star}(\bm{q})\|_{\mathcal{A}}^{\star} \leq 1.
\end{align}
Solving (\ref{eq14}) remains challenging due to the infinite-dimensional search over $[0, 1)^{4}$. To address this, we use results from  trigonometric polynomial theory \cite{dumitrescu2007positive} to propose an semidefinite relaxation (SDR) using matrix inequalities. First, we define the sum-of-squares relaxation degrees $s^{\prime}, r^{\prime}, l^{\prime}, k^{\prime}$, and extend $\bm{q}$ by zero-padding. The SDR of problem (\ref{eq14}) is then:
\begin{align}\label{20}
	&\max_{\bm{q}, \bm{Q}\succeq 0}~ \langle \bm{q}, \bm{y} \rangle_{\mathbb{R}}-\frac{\sigma}{4}\|\bm{q}\|_{2}\nonumber\\
	&\begin{bmatrix}
		\bm{Q} & \hat{\bm{Q}}^{H} \\
		\hat{\bm{Q}} &  \bm{I}_{T\times T}, 
	\end{bmatrix}\succeq 0,\nonumber\\
	&\mathrm{Tr}\left(\bm{\Theta}_{l^{\prime}}\otimes\bm{\Theta}_{k^{\prime}}\otimes\bm{\Theta}_{s^{\prime}}\otimes\bm{\Theta}_{r^{\prime}}\bm{Q}\right) = \delta_{l^{\prime}, k^{\prime}, s^{\prime}, r^{\prime}},
\end{align}
where $\bm{\Theta}$ are Toeplitz matrices, and $\delta_{l^{\prime}, k^{\prime}, s^{\prime}, r^{\prime}}$ is the Dirac function. This convex problem can be solved with tools like CVX \cite{grant2014cvx}.The transmitted data can be obtained by solving the following optimization problem:
\begin{align}\label{12}
	\min_{\bm{g}_{k}, \forall k, p_{r}\bm{p}_{r}, \forall r}\sqrt{\sum_{j=1}^{L} \bigg([\bm{y}_{w}]_{j}-\sum_{k =0}^{K-1}\bm{a}(\hat{\bm{\tau}}_{k})^{H}\tilde{\bm{D}}_{j}\bm{g}_{k})\bigg)^{2}}.
\end{align}
In the following subsections, we introduce various compression matrices, each leading to a different $\mu$, which in turn impacts the required number of samples at the receiver. This translates to variations in implementation complexity. For instance, by fixing the number of time slots, one can reduce the number of required receiving antennas, and vice versa, since a smaller $\mu$ implies that fewer samples are needed to recover both the radar parameters and the transmitted signal.

 \begin{figure*}[t]
 	\centering
 	\mbox{
 		\subfigure[]{\includegraphics[width=0.42\textwidth]{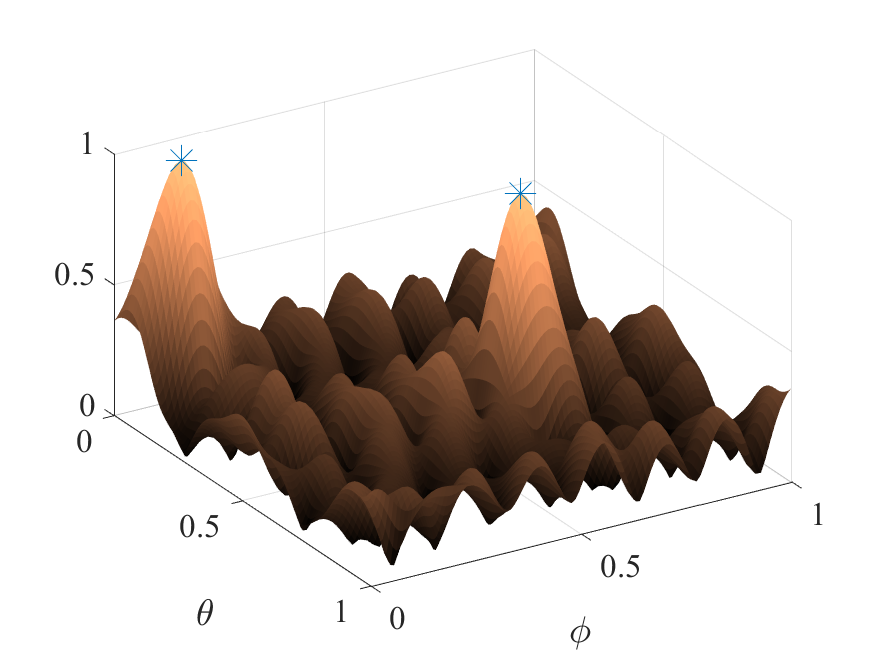}\label{fig.bound45}}
 		\subfigure[]{\includegraphics[width=0.42\textwidth]{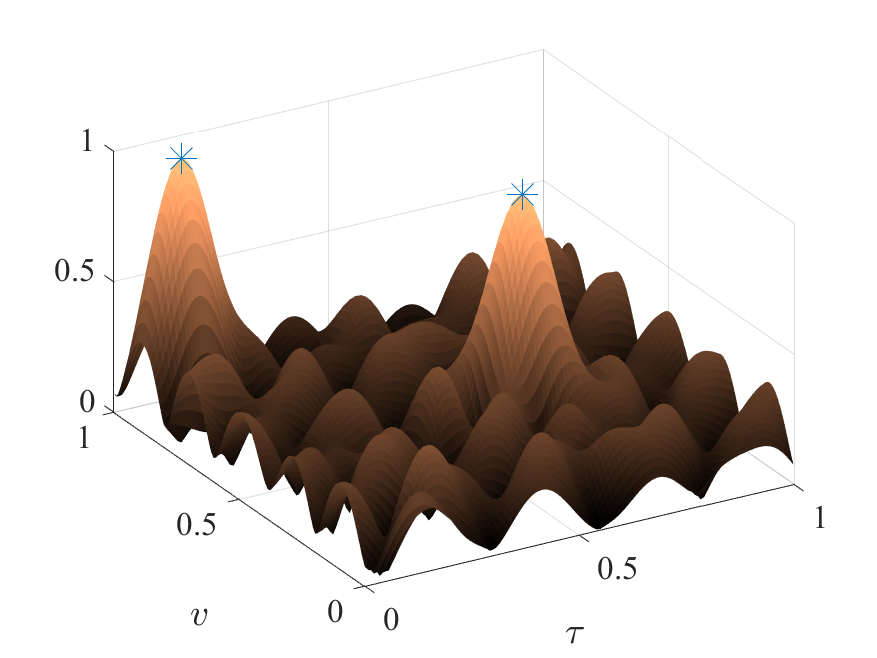}\label{fig.bound5}}
 	}
 	\caption{ The absolute value of the dual polynomial is shown in the $(\phi, \rho)$ and $(\tau, v)$ domains in Figs. \ref{fig.bound45} and \ref{fig.bound5}, respectively. Blue stars indicate the estimated radar parameters.} \label{fig.phase1}
 \end{figure*}
\subsection{Sensing Vectors with Independent Components.}
Suppose the components of $\bm{d}$ are independently distributed with zero mean and unit variance, which is isotropic. Moreover, if each component follows a light-tailed distribution, the corresponding measurements are clearly incoherent. A notable example is when $\bm{d} \sim \mathcal{N}(0, \bm{I})$, known as the \textit{Gaussian measurement ensemble}, which is one of the most frequently studied cases. In this situation, we can use $\mu = 6 \log T$ as previously discussed.

\subsection{Subsampled Orthogonal Transforms:}
Assume we have an orthogonal matrix such that $\bm{U}^*\bm{U} = n \mathbf{I}$ where $n$ is the size of the matrix. Now, consider a sampling process that selects rows of $\bm{U}$ uniformly and independently at random, then construct $\bm{D}$. When $\bm{U}$ is the DFT, this corresponds to the random frequency model described earlier. In this case, the distribution is isotropic, and $\mu = \max_{ij} |\bm{D}_{ij}|^2$. For cases where $\bm{D}$ is either a Hadamard matrix, we have $\mu = 1$.

\subsection{Subsampled Tight or Continuous Frames:}
We can extend the previous example by subsampling a tight frame or even a continuous frame. A key example of this is the Fourier transform with a continuous frequency spectrum, where 
\[
d(t) = e^{i 2 \pi \omega t}
\]
and $\omega$ is selected uniformly at random from $[0,1]$. This distribution is isotropic and satisfies $\mu = 1$. One real-world application of this concept is in magnetic resonance imaging (MRI), where frequency samples often do not align with an equispaced Nyquist grid. When time and frequency are swapped, this approach is aim to sampling a nearly sparse trigonometric polynomial at randomly chosen time points within the unit interval. In the following section, we carry out numerical experiments to evaluate the performance of the proposed estimator and compare the results for different compression matrices.

\begin{figure}[t]
	\centering
	\mbox{
		\subfigure[]{\includegraphics[width=0.25\textwidth]{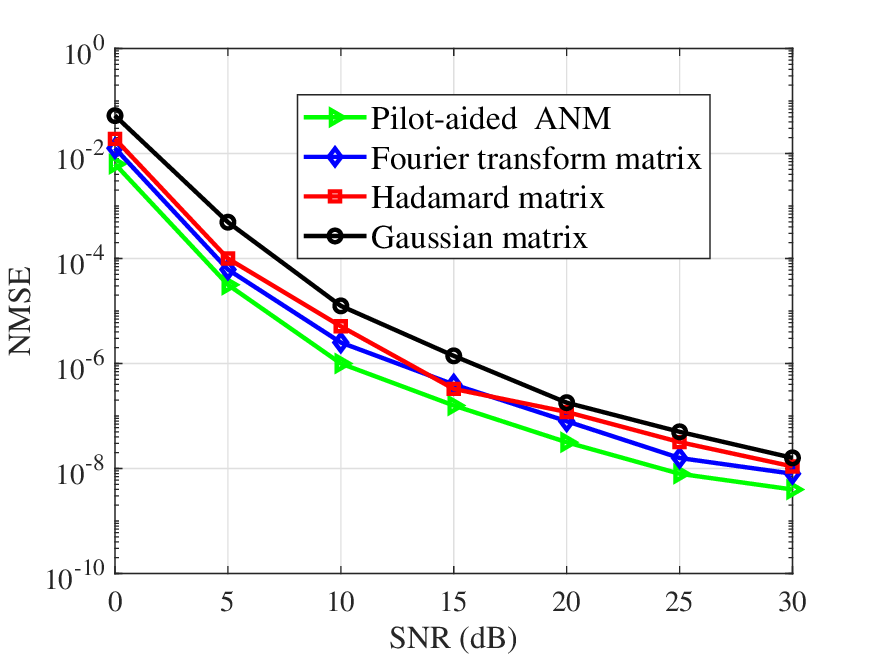}\label{fig.bound451}}
		\subfigure[]{\includegraphics[width=0.25\textwidth]{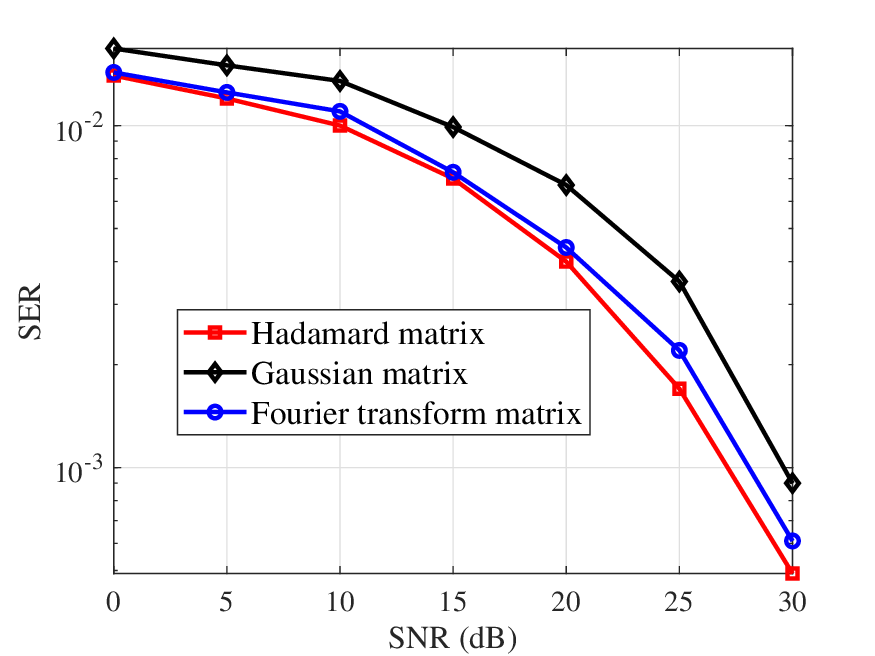}\label{fig.bound52}}
	}
	\caption{ The NMSE and SER of the proposed estimator, using different compression matrices, are compared with ANM as a benchmark in Figs. \ref{fig.bound451} and \ref{fig.bound52}, respectively. } \label{fig.phase2}
\end{figure}
\begin{figure*}[t]
	\centering
	\mbox{
		\subfigure[]{\includegraphics[width=0.3\textwidth]{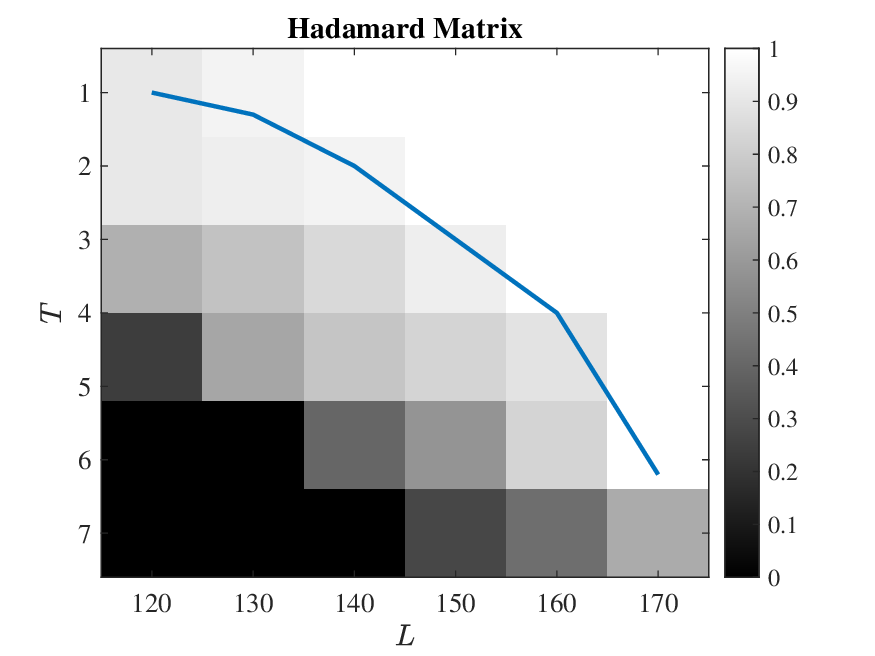}\label{fig.bound412}}
		\subfigure[]{\includegraphics[width=0.3\textwidth]{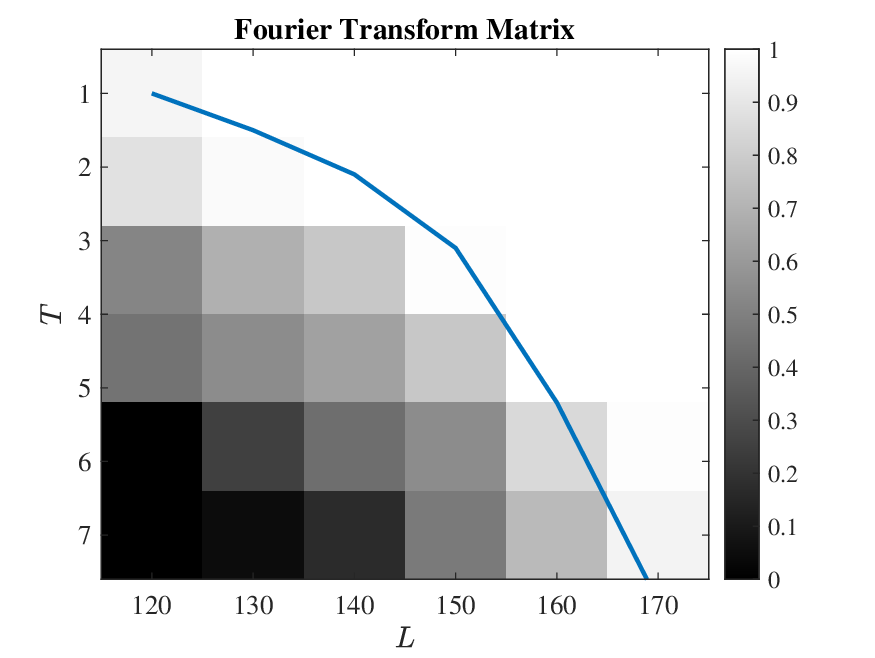}\label{fig.bound124}}
		\subfigure[]{\includegraphics[width=0.3\textwidth]{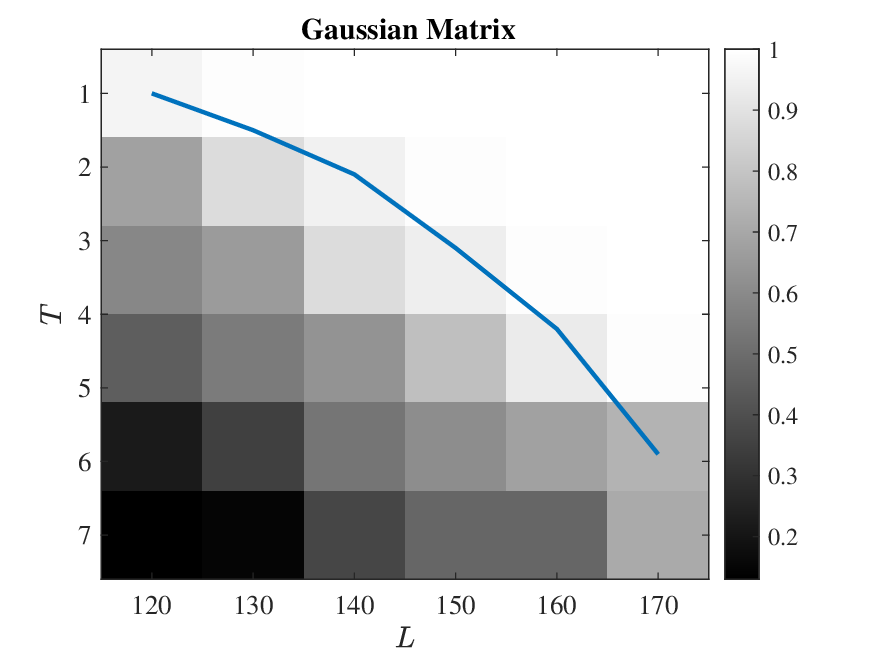}\label{fig.bound1245}}
		
	}
	\caption{The success rate of LANM versus the number of targets and the subspace dimension $T$ when the number of measurements changes in Figs. \ref{fig.bound412}, \ref{fig.bound412}, and \ref{fig.bound124} for Hadamard, Fourier transform, and Gaussian matrices, respectively, for $k=2$.} \label{fig.phase4}
		
	\vspace{-.5cm}
\end{figure*}

\section{NUMERICAL Results}
This section evaluates the performance of LANM for joint radar parameter and transmit data estimation, comparing it with ANM \cite{tang2020range}. Radar parameters are randomly generated, with target coefficients as i.i.d. zero-mean complex Gaussian. Number of observations and compression matrix size are provided in each simulation.  All ANM settings match LANM, except for the compression matrix.

Our evaluation starts by plotting the norm of the dual polynomial in $\mathcal{X}^{\star}(\bm{q}^{\star})$ where $\bm{q}^{\star}$ is the solution of (\ref{20}) with \(L = 225\) observed signals. We estimate radar parameters by identifying locations where the norm reaches 1. Fig. \ref{fig.phase1} illustrates the two-dimensional dual polynomial. Radar parameters \((\tau_k, v_k, \theta_k, \phi_k)\) are detected where the dual polynomial's magnitude is 1. For clarity, we show the dual polynomial in two-dimensional plots: one for \((\tau, v)\) with fixed \((\theta, \phi)\), and the other for \((\theta, \phi)\) with fixed \((\tau, v)\).

Before the next simulation, we need to first define the normalized mean-square error (NMSE)
as $\mathbb{E}[\|\bm{v}-\hat{\bm{v}}\|_{2}/\|\bm{v}\|_{2}]$, where $\bm{v}=\sum_{k=0}^{K-1}\alpha_{k}\bm{a}(\bm{\tau}_{k})^{H}$ and $\hat{\bm{v}}$ is the recovered radar quantity with the parameters $\hat{K}, \hat{\alpha}, \hat{\bm{\tau}}$. Next, we define the symbol error rate (SER) as $\mathrm{SER}=\mathbb{E}\big[\frac{1}{KT} \sum_{k=0}^{K-1}\sum_{t=1}^{T}\mathbb{I}_{(h_{k,t}-\hat{h}_{k,t})}\big]$
where $\mathbb{I}$ is a binary indicator such that $\mathbb{I}=0$  if $h_{k,t}=\hat{h}_{k,t}$, otherwise, $\mathbb{I}=1$.

We compare the NMSE and SER results of LANM with pilot-aided ANM and different compression matrices, including Fourier, Hadamard, and Gaussian, across varying SNR levels, defined as $\text{SNR}=10\log_{10}\frac{1}{\delta_{2}}$, with $L=225$ in in Figs. \ref{fig.bound451} and \ref{fig.bound52}, respectively. The results show that as SNR increases, the NMSE and SER improve for all methods and matrix types. Despite estimating both the channel and transmitted signal, LANM performs similarly to ANM, which relies on the known transmit signal.
With a fixed number of samples, both Hadamard and Fourier matrices achieve nearly the same NMSE and SER performance, as shown in Figs. \ref{fig.bound451} and \ref{fig.bound52}. However, the Gaussian matrix performs worse than the other two, aligning with the discussion in Section \ref{systems}, where $\mu=6\log T$ and larger than $\mu=1$ for the Hadamard and Fourier transform matrices. These findings also hold for SER, where Hadamard and Fourier matrices outperform the Gaussian matrix. Additionally, Fig. \ref{fig.bound52} highlights that as SNR increases, SER decreases, but performance declines with larger QAM constellations.

 We analyze the phase transition graphs of LANM, focusing on the average success rate as it relates to the number of targets and subspace dimension $T$ for different numbers of measurements and compression matrices. This analysis explores the trade-off between the number of detected symbols and radar channel recovery performance. In each of the $20$ Monte Carlo simulations, we compute the normalized error $|\bm{U} - \hat{\bm{U}}|{F}/|\bm{U}|{F}$, where $\hat{\bm{U}}$ represents the recovered lifted matrix. If the error is less than or equal to $10^{-3}$, the recovery is considered successful. 
 The graphs show that increasing the number of measurements leads to an improved number of detected symbols. A comparison of Figs. \ref{fig.bound412} and \ref{fig.bound124} indicates that the recovery performance of the Hadamard and Fourier transform matrices are almost the same and outperform the Gaussian matrix in Fig. \ref{fig.bound1245}, which is consistent with the discussion at the end of Section \ref{systems}. 
  The results across these matrices demonstrate similar trends in both NMSE and SER.

\vspace{-.1cm}

	\section{Conclusion}
In this paper, we proposed a LANM-based estimator for ISAC super-resolution receivers, enabling simultaneous radar target parameter estimation and communication symbol decoding. By employing different compression matrices—Fourier, Hadamard, and Gaussian—we demonstrated that LANM maintains robust performance across varying SNR levels, with similar trends in NMSE and SER across the matrices. The results highlight that increasing the number of measurements improves recovery performance. Our findings confirm LANM's ability to efficiently balance between sensing and communication tasks with flexibility in implementation complexity.
	
\vspace{-.6cm}

\bibliographystyle{IEEEtran}
	\bibliography{References}
\end{document}